\documentclass[aps,pra,twocolumn,amsmath,amssymb,showpacs]{revtex4}
\usepackage{graphicx}
\usepackage{dcolumn}
\usepackage{bm}
\usepackage{amsmath}
\input epsf

\begin{document}

\preprint{}

\title{Chemical pathways in ultracold reactions of SrF molecules}

\author{Edmund R. Meyer\footnotemark[1]}
\email{meyere@phys.ksu.edu}
\affiliation{JILA, NIST and University of Colorado, Department of Physics,
  Boulder, Colorado 80309-0440, USA}
\author{John L. Bohn}
\affiliation{JILA, NIST and University of Colorado, Department of Physics,
  Boulder, Colorado 80309-0440, USA}

\date{\today}

\begin{abstract}
We present a theoretical investigation of the chemical reaction 
SrF + SrF $\rightarrow$ products, focusing on reactions
at ultralow temperatures. We find that bond swapping, 
SrF + SrF $\rightarrow$ Sr$_2$ + F$_2$, is energetically forbidden at these 
temperatures. Rather, the only energetically allowed reaction is 
SrF + SrF $\rightarrow$ SrF$_2$ + Sr, and even then only singlet states of 
the SrF$_2$ trimer can form.  A calculation along a reduced reaction 
path demonstrates
that this abstraction reaction is barrierless, and proceeds
by one SrF molecule ``handing off'' a fluorine atom to the other molecule.
\end{abstract}


\maketitle

\footnotetext[1]{Now at Department of Physics, Kansas State University}

\section{Introduction}

After many years of experimental effort, chemical reaction dynamics
has now entered the cold and ultracold regime.  Robust techniques
such as buffer-gas cooling~\cite{Campbell09_book} and Stark 
deceleration~\cite{Meerakker09_book} 
have pushed the energy resolution of molecular beam techniques 
down to the mK level, resulting in new probes of chemical dynamics
\cite{Scharfenberg10_PCCP,Hummon10_preprint,ye10,Parazzoli10_preprint}.
Extending the low-energy limit even
farther, coherent optical techniques have produced samples of
alkali dimers in their absolute ground state, at temperatures
in the 1-100 $\mu$K range~\cite{rbcspaper,kkniscience,Danzl10_NatP,tokyoKRb}.
These ultracold molecules are so exquisitely sensitive
to comparatively weak influences, that chemistry can be studied
and controlled by exploiting quantum statistics \cite{Ospelkaus10_Science},
electric fields \cite{Ni10_Nature}, and confinement in optical lattices
 \cite{Gorshkov09_PRL,Quemener10_PRA,Micheli10_PRL,Miranda10_preprint}.

With these new capabilities naturally come questions of what can be
learned about chemical reaction dynamics under these novel
circumstances.  The {\it manipulation} of chemical reactions by external
electric and magnetic fields relies heavily on the 
behavior of long-range physics, where the molecules exert, say, dipolar
forces on one another, but are too far away
from one another to react. Indeed, the theoretical analysis of the reaction 
2KRb $\rightarrow$ K$_2$ + Rb$_2$, observed and studied at JILA,
took this point of view, by treating the actual reaction as a nearly
perfectly absorbing ``black box,'' which removed the molecules when 
they got close enough together, but without regard for what exactly happened to
them~\cite{iqbj10,ij10}.  

Vice versa, to {\it understand} more about chemical dynamics from these
experiments  would presumably require a scattering
simulation on a complete four-body potential energy surface (PES) for
the K-K-Rb-Rb system, which does not yet exist.  In very recent work,
however, Byrd {\it et al.} have described the main salient features of
this surface~\cite{byrd10}. First,  along its main reaction coordinate, the 
reaction presents no energetic barrier to reaction (consistent with the high 
reaction rates observed at JILA). Second, its transition state, at 
the borderline between reactants and products, represents a T-geometry 
characteristic of an insertion reaction, wherein a K atom from one 
molecule inserts itself between the K and Rb of the other.  This
circumstance suggests that the reaction proceeds by a complicated
four-body dance of the involved atoms, which may be revealed at ultralow
temperatures by characterizing the resonant states of the complex.  
Observing these transition-state resonances and interpreting their
influence in chemical reactions is a longstanding goal of
physical chemistry, and one in which the high energy resolution of
ultracold molecules may be a great help.

Set against this backdrop, it would also be useful to explore other
molecules reacting at ultralow temperatures, to gain additional insight into
what can be learned.  To this end, the SrF molecule is an appealing
candidate, and the one with which we deal in this article. SrF is a 
prime example of a class of molecules that have been identified as 
amenable to direct laser cooling from a beam~\cite{demille09}, and in fact 
laser cooling has been recently demonstrated~\cite{Shuman10_Nature}. It is 
quite polar (dipole moment $\approx 1.4$ Debye),
so that electric field manipulation is a possibility.  Moreover,
it has an open-shell $^2\Sigma$  ground state that
gives it a magnetic moment as well.  This circumstance opens
opportunities such as trapping the molecules magnetically, while
manipulating their interactions electrically~\cite{hudson07}.

In this article we explore the possibility and mechanisms for chemical reactions
of SrF molecules at ultralow temperatures.  We make several observations.
First, the exchange reaction
\begin{equation}\label{e:bs}
  2~{\rm SrF} \rightarrow {\rm Sr}_2 + {\rm F}_2 + \Delta E_{\rm ex}
\end{equation}
is energetically disallowed, as might be expected for a reaction that
turns two ionic bonds into two covalent ones.  Therefore, the reaction,
if it happens at all, must proceed by an abstraction reaction in which 
an atom jumps from one molecule to the other.  This is exactly the opposite
situation from what occurs in KRb, where the exchange is the only
possible reaction, and then only just barely~\cite{byrd10,hutson10,meyer10}

Second, we find that the Sr-abstraction reaction
\begin{eqnarray} 
    2 {\rm SrF} &\rightarrow 
    {\rm Sr}_2{\rm F} + {\rm F}
    + \Delta E _{\rm trimer}^{\prime} \label{e:trimer1} 
\end{eqnarray}
cannot occur at low temperature, whereas the F-abstraction
\begin{eqnarray}
    2 {\rm SrF} &\rightarrow 
    {\rm Sr}{\rm F}_2 + {\rm Sr}
    + \Delta E_{\rm trimer} \label{e:trimer2}
\end{eqnarray}
{\it can} occur, as it produces the deeply bound SrF$_2$ trimer.  
Moreover, this reaction, which is barrierless, can only occur in the 
singlet channel, i.e., only if the reaction takes place on the PES with 
total electronic spin $S=0$.  Therefore, chemical reactions are
expected to occur at quite high rates for unpolarized SrF molecules,
while they should be strongly suppressed for spin-polarized SrF,
which scatter primarily on the triplet surface.  Spin-rotation
couplings will ensure that this suppression is not complete~\cite{krems07}.
Nevertheless, one must be mindful of any opportunity to suppress
inelastic collisions, as they can easily destabilize the gas and
derail attempts to exploit the molecules for many-body physics
applications.

Third, we describe the basic physics of the abstraction process by 
looking at a restricted version of the four-body PES.  
We find that the F end of one SrF molecule approaches the Sr end 
of the other, as dictated by the dipole-dipole interaction between molecules.
Next, the F atom is handed off from one molecule to the other,
and the free Sr goes off by itself.  By this hand-off mechanism it is possible
that the reaction proceeds without forming a resonant complex, and
that its interpretation from ultracold collision data may be more
straightforward than the complete re-arrangement necessary in reactions
of KRb.  In any event, complementary dynamics should give us
complementary insights into these quite different reactions.

\section{{\it Ab initio} calculations}

In this section we will study the relevant molecular species:
the dimers SrF, Sr$_2$ and F$_2$; and the trimers
SrF$_2$ and Sr$_2$F.   All calculations of these species are performed
using the {\sc molpro} suite of {\it ab initio} electronic 
structure codes~\cite{molpro}. 
We use the relativistic effective core potential 
and associated basis set of the Stuttgart group for Sr~\cite{stutt} (ECP28MDF) and the 
augmented, correlation consistent valence triple (quadruple) zeta basis set (AVTZ 
(AVQZ)) of Dunning for F~\cite{dunning}. For the diatomic properties, the AVQZ basis 
is used.

We perform the calculations by first computing
a spin-restricted Hartree-Fock (RHF) wave function 
as a starting guess for a coupled-cluster singles, doubles, and 
non-iterative triples excitations calculation (CCSD(T))~\cite{ccsd1}.  The minimum 
energy configuration is obtained using the method of steepest descents. Properties 
such as dipole moment, polarizability, and quadrupole moment are calculated using 
the finite field approach. To compare the energies to the free atom limit, we use 
basis set superposition error (BSSE) corrections as given by the method of Boys and 
Bernardi~\cite{bsse}.
We calculate the vibrational frequencies by making 
the symmetry-independent displacements of the atoms and calculating the Hessian 
within the {\sc molpro} suite of routines. 

\subsection{Diatomic species}

In the collision of two SrF molecules there are three diatomic species to consider; 
the reactant SrF and the possible products Sr$_2$ and F$_2$. In the following we 
will investigate the relevant diatomic properties such as equilibrium bond length, 
well depth, and vibration constant. But first, an exercise in bond 
strengths can immediately inform us as to whether the reactions in Eq.~(\ref{e:bs}) 
can occur. On the left hand side of the reaction we have two ionic bonds. On the 
right hand side we have one van der Waals bond and one covalent bond; the former is 
very weak and the latter usually less bound than an ionic bond. An ionic bond can be 
thought of as a strengthening of the single covalent bond because of a transferring 
of charge from one atom to the other. From these simple considerations, we 
conclude that the reaction in (\ref{e:bs}) is energetically unfavorable at 
ultracold temperatures. 

Sr$_2$ is a van der Waals molecule. The ${}^1$S$_0$ atomic structure means that the 
the outer $s$-electrons are already paired up and therefore play a limited role in 
the bonding of two Sr atoms. Previous work has shown that this molecule is not 
deeply bound, with a well depth of $D_e=1081.8$~cm$^{-1}$~\cite{sr2,stein}. 
Because of the large binding length and relatively larger mass of Sr atoms, the 
vibrational constant is fairly small, $\omega_e=40.3$~cm$^{-1}$~\cite{stein}. 
Therefore, the energy required to dissociate this molecule from the zeroth 
vibrational level is given by $D_0=1061.6$~cm$^{-1}$.

F$_2$ is a covalently bonded species, with a ground state of 
${}^1\Sigma_{\rm g}^{\rm +}$ symmetry. F$_2$ has an appreciably large binding energy 
of $D_0=$12950~cm$^{-1}$~\cite{hao} and much smaller bond length compared 
to Sr$_2$. The unpaired $p$-electrons in the F atoms pair up and form a fairly deep 
well. Due to the lighter mass and tighter confining potential, the F$_2$ molecule has a 
comparatively large vibrational constant, 917~cm$^{-1}$~\cite{chen}. Therefore, the 
well depth, $D_e=13410$~cm$^{\rm -1}$ when combining the work of \cite{hao} and 
\cite{chen}.

SrF is a highly polar molecule. The willingness of F to take an extra electron, and 
of Sr to give one up, would lead one to conclude that at the minimum of the well the 
molecule is well described by a Sr$^{\rm +}$F$^{\rm -}$ configuration. We therefore 
compared the optimized geometry calculated at the RHF-CCSD(T)+BSSE calculation in two 
ways. The first was to compare the energy at the bottom of the potential energy surface 
to that of free Sr and F atoms. The other was to compare to free Sr$^{\rm +}$ and 
F$^{\rm -}$ ions and then calculate the ionization potential of Sr and the electron 
affinity of F. These calculations agreed with each other to within
several cm$^{-1}$. The well depth $D_e$ we report here refers to the energy required to
separate SrF into its neutral partners, Sr and F.

\begin{table}
  \caption{\label{t:diatoms} The molecular properties of Sr$_2$, F$_2$, and SrF. 
    Energies are in cm$^{\rm -1}$, bond lengths in $\AA$, and moments in atomic 
    units. Experimental values are given where known.}
  \begin{ruledtabular}
    \begin{tabular}{l|c|c|c|c|c|c}
      Molecule & $D_e$ & $R_e$ & $\omega_e$ & $d_m$ & $\alpha_z$ & $\theta_{zz}$\\
      \hline      SrF (this work)  & 44200 & 2.084 & 499 & 1.38 & 126 &  8.95\\
      ~~Expt. & 45290(560)\footnotemark[1] & 2.075\footnotemark[2] 
      & 502.4\footnotemark[2] & 1.36\footnotemark[3] & ~ & ~ \\
      ~~Theory & 45000\footnotemark[4] & 2.085 & 507 & ~ & ~ & ~\\
      \hline
      Sr$_2$ (this work) & 820 & 4.773 & 36.2 & ~ & ~ & ~\\
      ~~Expt.\footnotemark[5] & 1081.8 & 4.672 & 40.3 & ~ & ~ & ~\\
      \hline
      F$_2$ (this work) & 12880 & 1.410 & 927 & ~ & ~ & ~ \\
      ~~Expt. & 13410\footnotemark[6] & 1.411\footnotemark[7] & 917\footnotemark[7] 
      & ~ & ~ & ~ \\
      \hline      
    \end{tabular}
  \end{ruledtabular}
  \footnotetext[1]{Ref.~\cite{eng79}.}
  \footnotetext[2]{Ref.~\cite{herzberg}.}
  \footnotetext[3]{Ref.~\cite{ernst}.}
  \footnotetext[4]{Ref.~\cite{langhoff}.}
  \footnotetext[5]{Ref.~\cite{stein}.}
  \footnotetext[6]{Ref.~\cite{hao}, ZEKE/ion-pair imaging $D_0$ adjusted by $\omega_e$ of \cite{chen}.}
  \footnotetext[7]{Ref.~\cite{chen}.}
\end{table}

In Table~\ref{t:diatoms} we present the results of our RHF-CCSD(T)+BSSE 
calculations. All bond lengths ($R_e$) are in $\AA$, well depths ($D_e$) and 
vibration constants ($\omega_e$) are in cm$^{\rm -1}$, and associated dipole and 
quadrupole moments ($d_m$ and $\theta_{zz}$) are in atomic units 
($e~a_0$ and $e~a_0^2$, respectively). The corresponding experimental values are 
from the references given, and uncertainties are printed for those where the value 
was reported. The theoretical calculation of Langhoff {\it et al.}, is done at 
the configuration interaction with single and double excitations level of 
theory~\cite{langhoff}. Our current method yields a well depth smaller than 
the experimentally obtained value by $2\sigma$~\cite{eng79}. The bond length is 
in good agreement with that of ~\cite{langhoff}.
From these results we immediately see that the reaction in Eq.~(\ref{e:bs}) 
is energetically forbidden, by more than 70,000 cm$^{-1}$, and is of no
concern in an ultracold gas.

\subsection{Triatomic species}

There are two different triatomic species to consider as given in 
Eqs.~(\ref{e:trimer1}) and (\ref{e:trimer2}). 
At first glance, we might think both are quite easily allowed. From a simple bond 
strength argument, we can see that one is favored over the other. Particularly, that 
the triatom in Eq.~(\ref{e:trimer1}) is less bound than the triatom in 
Eq.~(\ref{e:trimer2}). The triatom in (\ref{e:trimer2}) has two ionic bonds, while 
the triatom in (\ref{e:trimer1}) can have only one. Therefore, from the fact that 
two ionic bonds should be more deeply bound than one, we expect the triatom in 
(\ref{e:trimer2}) to be more deeply bound than in (\ref{e:trimer1}). In addition, 
it is expected that the reaction in (\ref{e:trimer1}) is energetically unfavorable 
because there are stronger bonds on the LHS than on the RHS. However, the reaction 
in (\ref{e:trimer2}) cannot be determined from these arguments. Thus, it is to be 
calculated using the same {\it ab initio} methods as in the previous section. 
For the SrF$_2$ singlet state, we use the AVQZ basis for F as we did for SrF. However, 
for the remaining states we only used the AVTZ basis. This is because these states 
are so energetically forbidden (as we will see) that the less time-consuming calculation
is adequate for our purposes.

In Fig.~1, we give the three coordinates that are used in the 
optimization of the geometry. Using these three coordinates ($r_1$, $r_2$, and 
$\theta$), without constraining then, ensures that the electronic wave
function belongs to the C$_{\rm s}$ 
symmetry group. Thus, there are electronic states with even (A$^\prime$) and odd 
(A$^{\prime\prime}$) reflection symmetry through the plane containing the three 
atoms. 
When the two bond lengths are equal, the symmetry group is instead
expanded to C$_{\rm 2v}$. This added symmetry describes whether the molecule 
possesses even or odd reflection about the line which bisects the homonuclear 
bond. Where the ground state has equal bond lengths,  we have indicated this
with the appropriate notation; $A_1$ ($B_2$) have even reflection symmetry about 
the mid-point between the homonuclear bond and even (odd) reflection symmetry through 
the plane containing the three atoms. Similarly, $A_2$ ($B_1$) has odd reflection about 
the midpoint of the homonuclear bond with even (odd) reflection through the plane 
containing the three atoms.

\begin{figure}
  \label{f:triGeom}
  \begin{center}
    \resizebox{3.in}{!}{\includegraphics{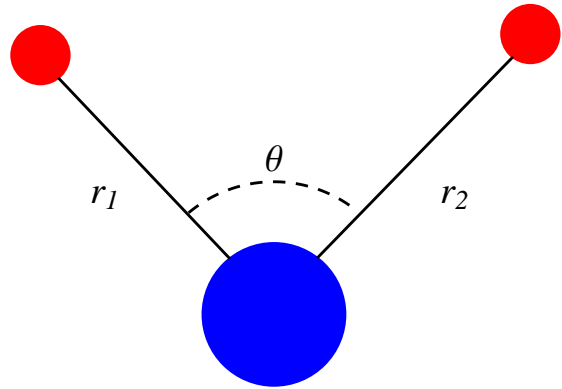}} 
  \end{center} \vspace{-.33in}
  \caption{(Color Online) Diagram of the geometries considered in the calculation 
    of the the molecules Sr$_2$F and SrF$_2$. The coordinates $r_1$, $r_2$, and 
    $\theta$ were varied and energy optimized to find the minimum energy 
    configuration in both the A$^\prime$ and A$^{\prime\prime}$ symmetry groups.}
\end{figure}

\begin{table}
  \caption{\label{t:triatom} The molecular properties of SrF$_2$ and Sr$_2$F. 
    Energies are in cm$^{\rm -1}$, bond lengths in $\AA$ and angles in degrees. 
    Previous theoretical values are given where applicable.}
  \begin{ruledtabular}
    \begin{tabular}{l|c|c|c|c|c}
      Molecule & Symm. & $D_e$ & $r_1$ & $r_2$ & $\theta$ \\
      \hline
      SrF$_2$ & ${}^1$A$_1$ & 91870 & 2.13 & 2.13 & 135.7 \\
      ~~Theory\footnotemark[1] & ~ & ~ & 2.16 & 2.16 & 138.8 \\
      ~~Expt.\footnotemark[2] & ~ & 89300 to 91900 & ~ & ~ & ~ \\
      \hline
      Sr$_2$F & ${}^2$B$_2$ & 52300 & 2.26 & 2.26 & 115.8\\
      \hline
      Sr$_2$F & ${}^2$B$_1$ & 43900 & 2.25 & 2.25 & 101.7\\
      \hline
      SrF$_2$ & ${}^3$A$^{\prime\prime}$ & 43400 & 2.10 & 3.72 & 70.4\\
      \hline
    \end{tabular}
  \end{ruledtabular}
  \footnotetext[1]{Ref.~\cite{kaupp91}.}
  \footnotetext[2]{Ref.~\cite{nistSrF2}. Extracted from enthalpy of formation.}
\end{table}

In Table~\ref{t:triatom} we present the results of the RHF-CCSD(T)+BSSE 
calculation. 
As is evident, the most deeply bound trimer is given by the ${}^1$A$_1$ 
symmetry for SrF$_2$. It is also the only one for which previous theoretical work 
is available for comparison. In Ref.~\cite{kaupp91}, Kaupp et al. present a 
theoretical study of many alkaline-earth metal atoms with two halogen atoms.
We note that alkaline-earth di-halides have historically been of significant interest
in chemistry, as some have bent geometries (hence possess permanent dipole moments
in their body frame) and some are linear (with no such dipole moment).  
A bend is not expected from classical models of charged particles interacting,
and therefore correctly predicting the bend is a measure of the
quality of the wave function. This is because the bend is a result of the presence 
of $d$-electron orbitals on the ground state. Our 
calculation agrees well with that of \cite{kaupp91} in the description of the bond 
length and bending angle. For the first time, we present the energy required to 
pull apart the molecule into its atomic constituents - atomization. This energy 
is greater than that of two SrF molecules and therefore, in the absence of a 
barrier, should react chemically at ultracold temperatures. 

Vice versa, the triplet SrF$_2$ molecule (last line of Table II) is far less bound, 
and indeed has binding energy similar to that of a single SrF molecule.  
A simple way to understand this is to think of this as a triplet covalent bond in 
F$_2$ with a Sr atom bound to one or the other F atoms ionically. The triplet bond in 
F$_2$ is much shallower than the singlet covalent bond in the ground state. Then the 
Sr comes along and binds to one of the F atoms. A simple charge population analysis 
shows that only the F atom closer to Sr captures charge. Importantly, the triplet trimer 
is energetically disallowed at ultracold (and even room) temperature studies of SrF 
collisions.

The other potential  trimer, Sr$_2$F, is reported  in the 
second and third lines of Table~\ref{t:triatom}. This molecule, while more deeply 
bound than the triplet state of SrF$_2$, is not deep enough to be chemically 
reactive, as expected from the simple bond arguments. Thus, this reaction 
will not proceed at ultracold temperatures. Notice that this molecule 
has a binding energy only slightly more than that of a single SrF diatom. The 
doublet nature of this molecule is due to the one unpaired electron that can be 
viewed as coming from the F atom plus a Sr$_2$ bond, or from the doublet 
nature of SrF plus a free Sr atom.

We therefore conclude that only the singlet electronic state of SrF$_2$ will 
form in the collision of two SrF molecules, at ultralow collision energy. 
We find that the energy released is $3470$~cm$^{\rm -1}$.
To be thorough, we must take into account the zero-point vibrational energy of the 
systems in consideration. In Table~\ref{t:diatoms} we gave the calculated and 
experimental values of the vibrational constants of SrF. We have calculated the 
vibrational constants of SrF$_2$ in the singlet electronic state: 
$\omega_{\rm sym}=472.9$, $\omega_{\rm asym}=478.7$, and 
$\omega_{\rm bend}=79.3$~cm${\rm -1}$, where these 
represent the symmetric (or breathing) mode, the anti-symmetric mode, and the 
bending mode of the triatomic molecule. This yields a zero-point energy of 
515.5~cm$^{\rm -1}$. Therefore, all the exoergic reactions are reduced by 
$\approx$~20~cm$^{\rm -1}$. It is evident that the added vibrational 
modes (three vs one in the diatom SrF) are not enough to overcome the energy 
released in the chemical reaction. In fact, there are roughly 150 vibrational 
states accessible in the reactants, given  using the calculated exoergicity of 
$\Delta_{\rm trimer} = 3450$~cm$^{\rm -1}$. 

To our knowledge, this quantity has been measured only once,
in a gas cell which maintained chemical equilibrium between
reactants and products at $\sim 1500$ K, using mass spectroscopy to
measure their relative abundance \cite{Hildenbrand68_JCP}.  
Knowing the temperature, this measurement provided information on the
enthalpy of formation $\Delta_f H$, which corresponds to the energy
released, $\Delta E_{\rm trimer}$ in Eq.~(\ref{e:trimer2}).
This measurement agrees that the reaction is exothermic, but by a more
modest value of $\Delta_f H = 2.1$ kcal/mol = $740$ cm$^{-1}$, than
we have calculated.  Given that the experimental uncertainty was
comparable to  700 cm$^{-1}$ and our computational uncertainty is perhaps
2000 cm$^{-1}$, the results are not too seriously in disagreement.  Ultimately, 
measurements in laser-cooled SrF samples should
sort out this issue in detail.

\section{Reaction path basics}

To compute the full PES of the four-body system involved in the reaction is
of course a complicated affair.  Here we instead compute selected
slices of this PES.  One is a reaction-path coordinate version that
will verify that the reaction is barrierless.  The second is a more detailed 
examination of the ``hand-off'' mechanism that drives this abstraction
reaction.  We will focus exclusively on the singlet PES, as it is the
only one leading to reactions in an ultralow temperature SrF gas. All 
calculations are performed with the ECP28MDF basis set and ECP of the Stuttgart 
group for Sr along with the AVTZ basis of Dunning for F. We use the 
RHF-CCSD(T)+BSSE level of theory.

A basic idea in constructing the reaction path surface is that the
relatively heavy Sr atoms move comparatively slowly, and therefore
the Sr-Sr distance can be regarded (approximately) as an adiabatic coordinate.
Fixing this distance, denoted $R_{\rm Sr-Sr}$, we optimize the 
coordinates $r_1$, $r_2$, $\theta_1$, and $\theta_2$ of the F atoms, as defined in
Figure 2, so as to minimize the energy.  For simplicity, we 
constrain all four atoms to lie in a plane. While this limits the possibility of 
the molecules changing the dihedral angle, it does not prevent all insertion 
type configurations from being explored. As we will see, the optimizations of the 
other coordinates suggest a preference for linear orientations upon approach. 

To track the progress of the reaction, we define the approximate reaction 
coordinate $\Delta$ via
\begin{equation}
  \label{e:reactcoord}
  \Delta = r_2 - r^\prime.
\end{equation}
In the limit that $\Delta$ is negative and large, it denotes the distance between
the two SrF diatoms, whereas when $\Delta$ is positive and large, it refers to
the distance between the SrF$_2$ trimer and the free Sr atom. 

\begin{figure}
  \label{f:react}
  \begin{center}
    \resizebox{3.25in}{!}{\includegraphics{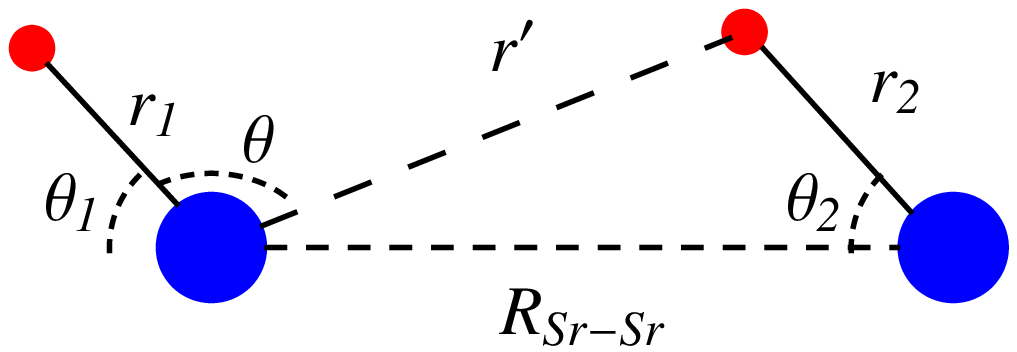}} 
    \resizebox{3.25in}{!}{\includegraphics{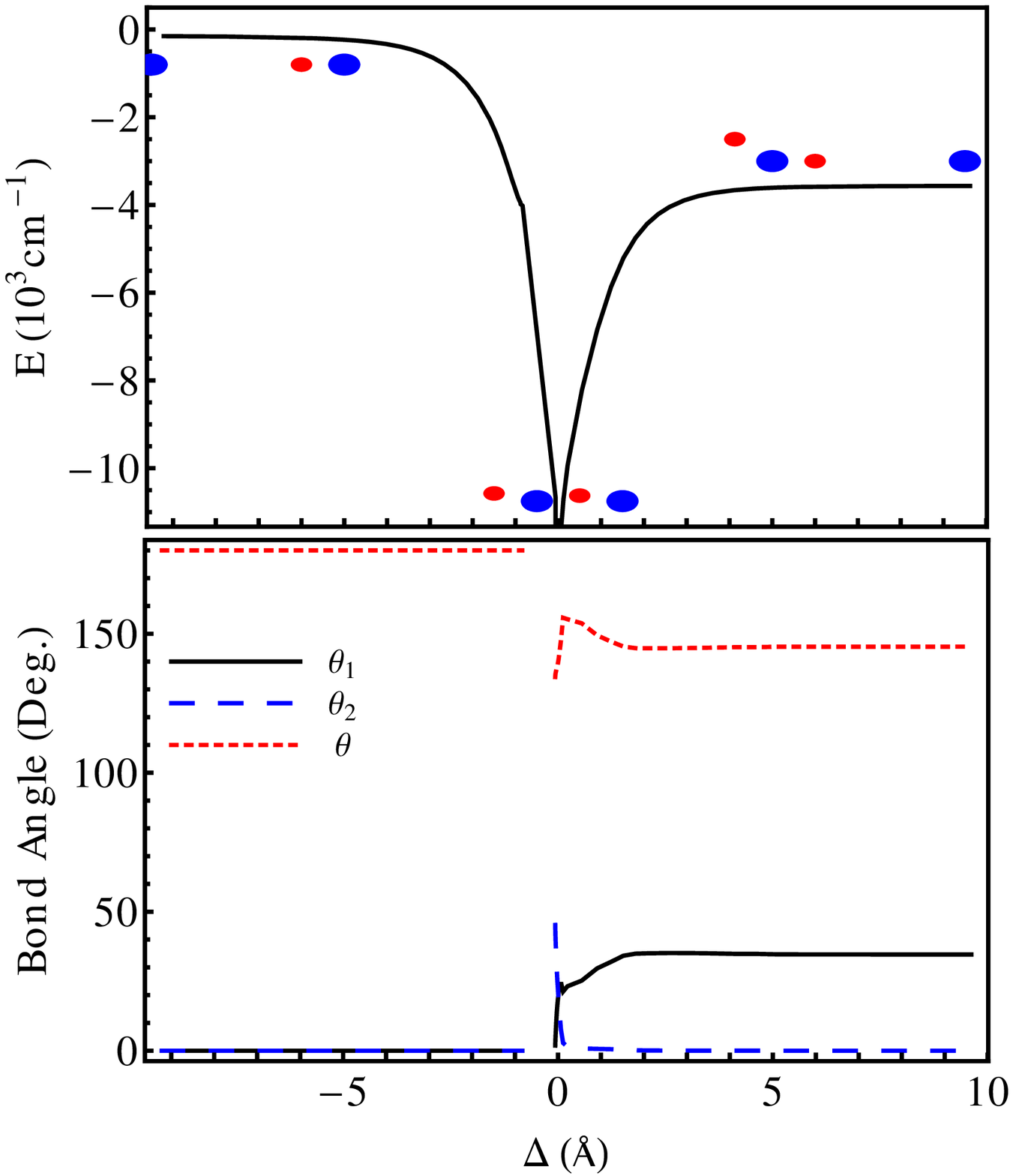}}
  \end{center} \vspace{-.33in}
  \caption{(Color Online) The top portion of the figure describes the geometry of the 
    situation. $r_1$, $r_2$, $\theta_1$, and $\theta_2$ are optimized so as to produce 
    the minimal energy for a fixed $R_{\rm Sr-Sr}$. The LHS describes the approach 
    of two dipole objects. On the RHS, the approach is more van der Waals like where 
    the polar SrF$_2$ polarizes the Sr atom. The middle is a sharp point that is an 
    artifact of the choice of $\Delta=r_2-r^\prime$. The larger, blue (smaller, red) 
    filled disks represent Sr (F) atoms. The approach to the transition state near 
    $\Delta=0$~$\AA$ is very nearly linear up to the point of F capture. Then $\theta_1$ 
    and $\theta_2$ rotate to form a complex which contains a slightly distorted SrF$_2$ 
    configuration. The lowest panel shows the optimized angles $\theta_1$ (solid, black 
    line), $\theta_2$ (large dashed, blue line), and $\theta$ (small dashed, red line). 
    The values remain fairly constant for $|\Delta|>1$~$\AA$, and change to form the 
    transition state in the region $|\Delta|<1$~$\AA$}
\end{figure}

The optimized energy is plotted as a function of 
$\Delta$ in the middle panel of Fig.~2, adjusted so that
the zero of energy refers to two free, separated SrF molecules.  
The PES along this path shows very little structure: it corresponds to almost
pure dipolar attraction between reactants for $\Delta <0$,
and almost pure van der Waals attraction between the products for $\Delta >0$.
In particular, this cut through the PES shows clearly that there is {\it no barrier}
to reaction, at least along the minimum energy path.

This choice of reaction coordinate is useful, but also leads to a structural 
discontinuity in the surface of Fig.~2. This discontinuity is made clear in 
the last panel of Fig.~2 which depicts the variation in the angles 
$\theta_1$, $\theta_2$, and $\theta$ (see top panel of Fig.~2). This is an 
artifact of the choice in approximate reaction coordinate. Physically, this 
discontinuity arises because $R_{\rm Sr-Sr}$ is not strictly an adiabatic coordinate. 
At some point, the intermediate F atom glides from its local minimum (attached to 
the right-hand Sr atom) to its global minimum (attached to the SrF on the left). When
this occurs, a single value of $R_{\rm Sr-Sr}$ corresponds to two distinct
values of the reaction coordinate $\Delta$ where the adiabatic energy
is the same.  This mechanism will be clarified by looking at the PES
from a different perspective, below.

For $\Delta<0$, the system is described by a linear geometry 
($\theta_1=\theta_2=0^\circ$). As the diatoms approach one another, they reach a 
location in $\Delta$ where the discontinuity occurs. From the $\Delta>0$ side, the 
values of $\theta_1$, $\theta_2$, and $\theta$ are fairly constant. As $\Delta$ is 
decreased toward zero by decreasing R$_{\rm Sr-Sr}$, the 4-atom system reaches a minimum 
near $\Delta=0$. As the R$_{\rm Sr-Sr}$ is decreased further, $\theta_1$ approaches 
$0^\circ$ and the system approaches the geometrical discontinuity from the other side. 
This leads to a fairly smooth reaction path in the coordinate $\Delta$, but hides the 
change in the geometrical configuration of the 4-atom system.

Using the information contained in the values of $\theta_1$, $\theta_2$, and $\theta$ we 
are led to study the F capture process by analyzing the linear configuration. Even 
though the final trimer's ground state is bent, the energy of the linear configuration 
is only $\sim$500~cm$^{-1}$ above the bent configuration of SrF$_2$. For this reason, we 
can make a reasonable qualitative description of the reaction by constraining all atoms 
to lie on a line, as in the upper panel of Fig.~3.

We again treat the Sr-Sr distance as an adiabatic coordinate.
Then, for fixed  Sr-Sr distances $R_{\rm Sr-Sr}$, we  vary 
the location of the middle F atom, describing it with coordinate $r^{\prime}$
as depicted at the top of Fig.~3. The leftmost F atom is fixed to the 
diatom bond length of SrF, $r_1$. Because the bond length in SrF 
and SrF$_2$ are so similar, this is a reasonable assumption. Using the 
RHF-CCSD(T)+BSSE method we calculate the potential experienced by the middle F 
atom. The results are presented in Fig.~3 for
various values of $R_{\rm Sr-Sr}$. The BSSE compares the 
energy to the atomization limit. In the figure, we offset the energy by the 
binding energy of two SrF molecules. Thus, zero energy is the energy of 
two SrF diatoms.

\begin{figure}
  \label{f:handoff}
  \begin{center}
    \resizebox{3.25in}{!}{\includegraphics{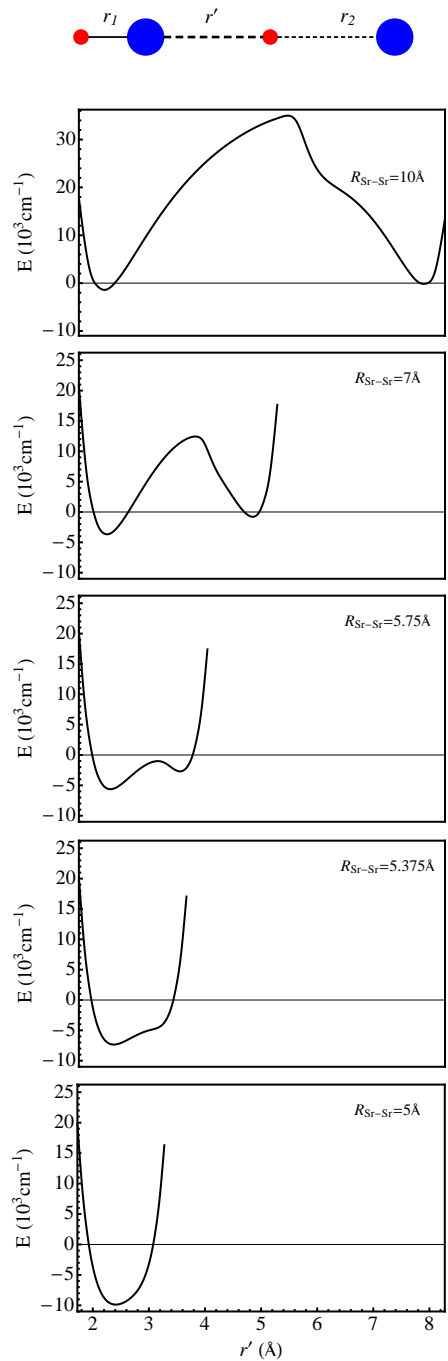}} 
  \end{center} \vspace{-.33in}
  \caption{(Color Online) The top portion gives the geometry of the system. The 
    bond length $r_1$ is fixed while $r_2$ is varied. $r^\prime$ gives the distance to 
    the middle F to the other Sr atom and is called the hand-off coordinate since it 
    describes the handing off of a F to one Sr from the other Sr. Each subsequent plot 
    is for a fixed value of $R_{\rm Sr-Sr}$ given in the upper RHS. As the value of 
    $R_{\rm Sr-Sr}$ is reduced, the barrier to forming the system F-Sr-F + Sr is 
    diminished until it no longer stops the transfer of F. The calculations were 
    performed at the RHF-CCSD(T)+BSSE with the Sr ECP28MDF and F AVTZ bases 
    respectively.}
\end{figure}

In the first plot of Fig.~3 the two Sr atoms are far from one another.
The potential for the F atom has two minima, representing the fluorine
attached to the original Sr atom (right hand well) or else attached to the
other SrF to form the product SrF$_2$ (left hand well). The latter well being 
deeper signifies that the products SrF$_2$ + Sr are energetically favored and the
reaction is exoergic, even within a strictly linear geometry.  Between these minima
stands a high barrier, which naturally prevents the F atoms from
jumping from reactants to products when the Sr atoms are this far apart.

In subsequent panels of Fig.~3, the two Sr atoms approach one another.
In each case the barrier lowers until it eventually disappears altogether, and the
F atom rests at the bottom of a single minimum in the last plot of Fig.~3.
This configuration defines the four-body transition state with $\Delta \approx 0$~$\AA$.
At this point the Sr atom can recede, and to the extent that its motion really is
adiabatic, the F atom will tend to remain in the lower well, thus
finding itself a part of the SrF$_2$ final product.  We say that this F-abstraction 
reaction occurs by a ``hand-off'' mechanism, whereby the right-hand Sr atom 
approaches, gently hands off the F, and then goes away.  As compared to
the KRb-KRb surface \cite{byrd10}, this reaction is less likely to partake in a 
complicated dance of the four atoms, becoming thoroughly enmeshed in a four-body 
transition-state complex.

\section{Conclusions}

In summary, we have established that ground-state SrF molecules will
indeed by chemically reactive, even at ultralow temperatures that will
be achieved via laser cooling.  The only possible outcome of such a
reactive collision would be the singlet trimer SrF$_2$, plus a free Sr
atom.  Because of the need to produce a singlet final state, chemical
reactivity should be strongly suppressed in a spin-polarized sample.
This species therefore looks like a promising candidate for ultracold
molecular studies.  
On the one hand, SrF may serve
as a useful platform for probing and understanding abstraction
reactions in unprecedented detail.
On the other hand,  if producing a spin-polarized sample suppresses
reactions sufficiently,  there is also hope that the molecules may live
long enough to perform interesting experiments on dipolar degenerate
quantum gases.  To estimate if this is so, future work will need
to study in more detail both the triplet surface, and its detailed
coupling to the reactive singlet surface.

\acknowledgments
We are grateful for support from the NSF. We acknowledge useful discussions with 
G. Qu{\'e}men{\'e}r about chemical reaction processes.

\end{document}